\documentclass[aps,superscriptaddress,groupedaddress,prb,twocolumn]{revtex4}%
\usepackage{amsfonts}
\usepackage{amsmath}
\usepackage{amssymb}
\usepackage{graphicx}%

\newcommand{\ket}[1]{| #1 \rangle}
\newcommand{\bra}[1]{\langle #1 |}

\newcommand{\beq}{\begin{eqnarray}}
\newcommand{\eeq}{\end{eqnarray}}

\begin{document}
\title{Vibrationally-mediated molecular transistors}
\author{D. H. Santamore}
\affiliation{Department of Physics, Temple University, Philadelphia, PA 19122, USA}
\affiliation{Advanced Science Institute, RIKEN, Saitama 351-0198, Japan}
\author{Neill Lambert}
\affiliation{Advanced Science Institute, RIKEN, Saitama 351-0198, Japan}
\author{Franco Nori}
\affiliation{Advanced Science Institute, RIKEN, Saitama 351-0198, Japan}
\affiliation{Physics Department, University of Michigan, Ann Arbor, Michigan, 48109, USA}

\begin{abstract}
We investigate the steady-state electronic transport through a suspended dimer
molecule coupled to leads. When strongly coupled to a vibrational mode, the electron
transport is enhanced at the phonon resonant frequency and higher-order resonances.
The temperature and bias determines the nature of the phonon-assisted resonances, with clear absorption and emission
peaks. The strong coupling also
induces a Frank-Condon-like blockade, suppressing the current between the resonances.
We compare an analytical polaron transformation method to two exact numerical methods: the Hierarchy equations of motion and an exact diagonalization in the Fock basis.  In the steady--state, our two numerical results are an exact match and qualitatively reflect the main features of the polaron treatment. Our results indicate the possibility of a new type of molecular transistor or sensor where the current can be extremely sensitive to small changes in the energies of the
electronic states in the dimer.
\end{abstract}


\date{\today }
\maketitle

\section{Introduction}

Electron transport through real, and artificial, suspended molecular systems
has revealed a rich tapestry of physical
effects\cite{gen1,gen2,gen3,gen4,gen5,gen6,gen7,gen8}. Nano-mechanical systems
offer the promise of observing quantum effects in massive objects
\cite{sense1,sense2,sense3}, which can be cooled, driven, measured and
manipulated using electron transport
\cite{massetti,lambert1,lambert2,gen7,koch2,koch3,armour,keith,konrad,girvin,deb}.
Single molecule electronics offers similar promises, with the benefit of
stronger coupling of vibrational and oscillatory modes to the transport
process, and a large range of practical
applications\cite{app1,app2,app3,app4,app5,app6,app7,app8,app9,app10,app11,app12,fujita}%
. Current efforts on single-molecule electronics include the exploration of
the electronic level structure and its effect on electronic transport together
with the development of electronic devices and applications\cite{BookME}.

Since the performance of molecular electronic devices depends on electron
transfer between molecules, it ultimately requires a good understanding of
coherent transport and dynamics. Molecular electronics has a variety of
predecessors in chemistry and chemical reactions, known as electron-transfer
reactions \cite{MR87,many03}. One particular example, the
donor-bridge-acceptor (DBA) system has been studied for a long time and used
to probe the mechanisms of charge transfer\cite{BFetc08}. In this system, the
donor molecule donates an electron to the acceptor via the third non-rigid or
rigid molecule called a \textquotedblleft bridge\textquotedblright\ (i.e.,
redox reaction). In the lab it is possible to construct a molecular
electronics device by attaching electrodes to the donor and acceptor molecules
of a mixed-valence dimer and apply a bias voltage to cause electron transport. Aviram
and Ratner\cite{AR74} made the first metal-DBA-metal system in analogy to p-n
junctions. The transfer rate of incoherent transfer can be explained within the
well-known F\"{o}rster and Dexter theory\cite{F59,D53,JJS02,JNS04}, and until
recently, coherent transport had been regarded to play only a small role in charge-transfer dynamics of molecules. However, this view has been changed in the
last decade and some experiments indicate that coherent transport is indeed
important\cite{sholes,jang}.

In a slightly different context, recent experiments in photosynthetic
complexes have shown that \emph{excitonic} coherence lasts surprisingly long
at non-zero temperature: up to 300 fs at room
temperature\cite{Engeletc07,Petc10,Cetc10}. It has recently been
proposed\cite{mancal,mancal2} that the vibrational modes in the protein environment
may also play a role in the observed oscillations, long coherence time, and
high transport efficiency. Therefore, an understanding of the intricacies of
both coherent electron transport and coherent exciton transport transport in
molecules, coupled to vibrational modes, is vital
\cite{BFetc08,MLetc03,HMK05,CFAH07,YAYSS02,GB04,OMetc08,Engeletc07,GM06}.

In this paper, we examine a model of electron transport through a two-site
system coupled linearly to a single vibrational mode. Motivated by recent
progress in single-molecule experiments, we study the case when the electron-phonon
coupling can be strong (non-perturbative), and where the phonons can have a life-time much longer than the
electronic transport process. Throughout we use the language of a single
mixed-valence dimer molecule coupled to electronic leads. One electrode (source) is
attached to one site of a dimer molecule (donor) and another electrode (drain)
is attached to the other site (acceptor). This system is analogous to the DBA
system, but instead of incoherent transport, we include the possibility of
coherent electron transport and analyze steady-state dynamics and current noise power spectra.

We begin with a semi-analytical treatment based on a polaron-transform,
analogous to a single-mode version of the traditional \textquotedblleft
non-interacting blip approximation\textquotedblright(NIBA)
\cite{LeggettRMP,Dekker86}, which can give us some insight on a regime not
easily accessible with numerical methods (non-equilibrium transport, un-damped vibrational mode and strong
coupling). Such an approach has been used
elsewhere to study the transport through single and double quantum
dots\cite{Brandes05, lambert1, lambert2,esposito}, photosynthetic
complexes\cite{bremer} (and similar, though slightly different, techniques
were used in \cite{YCC,mancal,mancal2}), and molecules. Our derivation follows in the
same vein as these earlier works. We follow this with two exact numerical
treatments, one based on the Hierarchical equations of motion
\cite{ma,tanimura} with a Lorentz bath, the other an exact diagonalization of
the Hamiltonian in the Fock basis.

Using these various methods we show that in certain conditions the vibrational
mode coupling can strongly enhance electron transport in a selective manner,
suggesting the possibility of using vibrationally-assisted electronic
transitions as a highly sensitive transistor.

\section{Hamiltonian}

Consider a dimer molecule (denoted below by sites left (L) and right (R)) with
each site coupled to a lead (fermionic reservoir) so that the dimer functions
as a bridge. A vibrational mode, e.g., a flexural or dilational mode of the
molecule (or bridge in the D-B-A system), is coupled to the electrons in the left
and right sites of the dimer. The idea of strong coupling to one vibrational
mode is not far-fetched\cite{app12,Brandes05}: Such coupling has been shown
experimentally in some nano-electronics
devices\cite{Stetten10,carbon1,carbon2, carbon3,carbon4}.

The Hamiltonian of this system, which is analogous to the Hubbard-Peierls model
in the large on-site repulsion limit and with spin suppressed, can be written
as
\begin{equation}
H=H_{0}+H_{T}+H_{V}\nonumber
\end{equation}
where%
\begin{align}
H_{0}  &  =H_{\mathrm{mol}}+H_{\mathrm{osc}}+H_{\mathrm{leads}}\nonumber\\
H_{T}  &  =H_{\mathrm{tunn}}+H_{\mathrm{couple}}\\
H_{V}  &  =V_{L}a_{L}^{\dag}d_{S}+V_{R}d_{D}^{\dag}a_{R}+V_{L}d_{S}^{\dag
}a_{L}+V_{R}a_{R}^{\dag}d_{D}%
\end{align}%
\begin{align}
H_{\mathrm{mol}}  &  =\frac{1}{2}\varepsilon_{L}a_{L}^{\dag}a_{L}+\frac{1}%
{2}\varepsilon_{R}a_{R}^{\dag}a_{R}\\
H_{\mathrm{osc}}  &  =\hbar\omega_{m}b^{\dag}b\\
H_{\mathrm{leads}}  &  =\sum_{k}\left(  \hbar\omega_{k}^{S}d_{s,k}^{\dag}d_{s,k}%
+\hbar\omega_{k}^{D}d_{D,k}^{\dag}d_{D,k}\right) \\
H_{\mathrm{tunn}}  &  =T_{c}\left(  a_{L}^{\dag}a_{R}+a_{R}^{\dag}a_{L}\right) \\
H_{\mathrm{couple}}  &  =g_{L}\left(  b^{\dag}+b\right)  a_{L}^{\dag}a_{L}-g_{R}\left(
b^{\dag}+b\right)  a_{R}^{\dag}a_{R}.%
\end{align}
Here $H_{0}$\ is the free Hamiltonian of the dimer $H_{\mathrm{mol}}$, the oscillator $H_{\mathrm{osc}}$, and the
source and drain reservoirs $H_{\mathrm{leads}}$ attached to the left (right) sites, with resonant
frequencies $\omega_{L}$, $\omega_{R}$, $\omega_{m}$, $\omega_{k}^{S}$, and
$\omega_{k}^{D}$, respectively. Also, $T_{c}$ is the transmission coefficient,
and $g_{L\left(  R\right)
}$\ is the electron-vibrational-mode coupling coefficient. The annihilation
(creation) operators for the electron in the molecules are $a_{L\left(
R\right)  }\left(  a_{L\left(  R\right)  }^{\dag}\right)  $, for the lead
reservoirs are $a_{S\left(  D\right)  }\left(  a_{S\left(  D\right)  }^{\dag
}\right)  $, and for the oscillator is $b\left(  b^{\dag}\right)  $. The
interaction Hamiltonian has three parts: lead electrode-molecule coupling
$H_{V}$, electron tunneling between the two sites in the dimer molecule
$H_{\mathrm{tunn}}$, and the linear coupling between the dimer and the vibrational mode
$H_{\mathrm{couple}}$. The coupling between the electrons and the vibrational mode
effectively modify the frequency of the electrons in the left and right sites
of the molecule as $\omega_{L\left(  R\right)  }\rightarrow\omega_{L\left(
R\right)  }^{\mathrm{eff}}$, where $\omega_{L\left(  R\right)  }^{\mathrm{eff}%
}=\omega_{L\left(  R\right)  }+g_{L\left(  R\right)  }\left(  b^{\dag
}+b\right)  $, while the tunneling barrier is unaffected by the coupling.

Here we adopt the notation, for brevity,
\begin{equation}
s\equiv a_{L}^{\dag}a_{R}\quad s^{\dag}\equiv a_{R}^{\dag}a_{L}%
\end{equation}
In the limit of large Coulomb repulsion to double occupation (a term not explicitly included in the Hamiltonian for simplicity), only a single electron occupies the entire molecule dimer at once.
For the moment, we retain the
notation for the general case, but will soon explicitly use this assumption.

The above Hamiltonian is the most general and exact form. To find
quasi-analytical results, this Hamiltonian can be transformed using the polaron
transformation to eliminate the interaction term, but which in turn modifies
the tunneling term $H_{T}$.
\begin{equation}
H_{T}=T_{c}\left(  s^{\dag}X^{\dag}+sX\right)
\end{equation}
where%
\begin{align}
X  &  \equiv\exp\left[  z\left(  \hat{b}-\hat{b}^{\dag}\right)  \right] \\
X^{\dag}  &  \equiv\exp\left[  -z\left(  \hat{b}-\hat{b}^{\dag}\right)
\right]
\end{align}
with $z=g/\omega_{m}$, $g=g_{L}=g_{R}$. Note that we have ignored a constant
energy shift induced by this transformation.

The dynamics of the system can be analyzed through the master equation, where the
coherent electron transport through the molecules is associated with the
off-diagonal operators $s$, $s^{\dag}$. In the next section we explicitly give details of the
master equation.

\section{Master equation}

To obtain the master equation, we switch to the interaction picture. In this
picture $a_{L}^{\dag}a_{L}$ and $a_{R}^{\dag}a_{R}$ are unchanged since they
commute with the non-interacting Hamiltonian:
\begin{align}
a_{L}^{\dag}a_{L}\left(  t\right)   &  =n_{L}\left(  t\right)  =n_{L}\\
a_{R}^{\dag}a_{R}\left(  t\right)   &  =n_{R}\left(  t\right)  =n_{R} .%
\end{align}
For the term $H_{T}$, we define new operators that group the electron
tunneling terms and vibrational mode operators:%
\begin{align}
\tilde{s}\left(  t\right)   &  =p\exp(i\epsilon t)X_{t},\\
\tilde{s}^{\dag}\left(  t\right)   &  =p^{\dag}\exp(-i\epsilon t)X_{t}^{\dag},
\end{align}
where $\epsilon\equiv\epsilon_{L}-\epsilon_{R}$. Then%
\begin{equation}
H_{T}\left(  t\right)  =T_{c}\left[  \tilde{s}\left(  t\right)  +\tilde
{s}^{\dag}\left(  t\right)  \right]  \label{H_T}%
\end{equation}

In the interaction picture, the equation of motion is%
\begin{align}
\frac{d\rho\left(  t\right)  }{dt} &  =-\frac{i}{\hbar}\left[  H_{\mathrm{int}}\left(
t\right)  ,\rho\left(  t\right)  \right]  \nonumber\\
&  =-\frac{i}{\hbar}\left[  H_{T}\left(  t\right)  +H_{V}\left(  t\right)
,\rho\left(  t\right)  \right]  ,
\end{align}
and the density matrix can be obtained by iterations. By keeping terms to
second order in the couplings to the the source and drain, and making the
standard Born-Markov approximation, the density matrix $\rho\left(  t\right)
$ can be approximated as $\rho\left(  t\right)  \simeq\rho_{0}^{DO}\left(
t\right)  \otimes\rho_{0}^{\mathrm{res}}$. Then, after tracing out the leads, the
master equation can be written in the Lindblad form%
\begin{align}
\frac{d\rho^{DO}\left(  t\right)  }{dt} &  =-\frac{i}{\hbar}\left[
H_{T}\left(  t\right)  ,\rho^{DO}\right]  \nonumber\\
&  +\frac{\Gamma_{L}}{2}\left(  1-f_{S}\right)  \mathcal{D}\left[  a_{L}^{\dag}\right]
+\frac{\Gamma_{L}}{2}f_{S}\mathcal{D}\left[  a_{L}\right]  \nonumber\\
&  +\frac{\Gamma_{R}}{2}\left(  1-f_{D}\right)  \mathcal{D}\left[  a_{R}^{\dag}\right]
+\frac{\Gamma_{R}}{2}f_{D}\mathcal{D}\left[  a_{R}\right] ,
\end{align}
where%
\begin{align}
\mathcal{D}\left[  a_{i}^{\dag}\right]   &  \equiv a_{i}^{\dag}a_{i}\rho
^{DO}\left(  t\right)  -2a_{i}\rho^{DO}\left(  t\right)  a_{i}^{\dag}%
+\rho^{DO}\left(  t\right)  a_{i}^{\dag}a_{i}\label{D}\\
\mathcal{D}\left[  a_{i}\right]   &  \equiv a_{i}a_{i}^{\dag}\rho^{DO}\left(
t\right)  -2a_{i}^{\dag}\rho^{DO}\left(  t\right)  a_{i}+\rho^{DO}\left(
t\right)  a_{i}a_{i}^{\dag}\label{D2}%
\end{align}
and $f_{S}$\ and $f_{D}$ are the Fermi distribution functions for the left and
right reservoirs, respectively, $f_{S}=\left[  e^{\hbar\beta\left(
\epsilon_{L}-\mu_{L}\right)  }+1\right]  ^{-1}$, and $f_{D}=\left[  e^{\hbar
\beta\left(  \epsilon_{R}-\mu_{R}\right)  }+1\right]  ^{-1}$.  In all the later
results we choose an infinite bias so that $f_{s}=1$, $f_{d}=0$.
Since we are interested in the strong-coupling limit, one may also argue that the
perturbative coupling to the leads should be derived after diagonalization of the
combined electronic/vibrational system. This is an interesting avenue for
future work.

\section{Expectation values}

At this point the molecule-vibrational mode coupling is still described
exactly. To achieve an analytical solution it is convenient to rewrite the
master equation as a closed set of equations for the expectation values of the
various molecule operators, following the formulation used by
Brandes\cite{Brandes05} to describe quantum dots coupled to a bath of
oscillators. To do so we need the following commutators,
\begin{align}
\left[  n_{L}\left(  t\right)  ,H_{T}\left(  t^{\prime}\right)  \right]   &
=T_{c}\left[  \tilde{s}\left(  t\right)  -\tilde{s}^{\dag}\left(  t\right)
\right]  \label{L}\\
\left[  \tilde{s}\left(  t\right)  ,H_{T}\left(  t^{\prime}\right)  \right]
&  =T_{c}e^{i\epsilon\left(  t-t^{\prime}\right)  }\left\{  n_{L}X_{t^{\prime
}}^{\dag}X_{t}-n_{R}X_{t^{\prime}}^{\dag}X_{t}\right\}  \\
\left[  \tilde{s}^{\dag}\left(  t\right)  ,H_{T}\left(  t^{\prime}\right)
\right]   &  =T_{c}e^{i\epsilon\left(  t-t^{\prime}\right)  }\left\{
n_{R}X_{t^{\prime}}X_{t}^{\dag}-n_{L}X_{t^{\prime}}X_{t}^{\dag}\right\}
\end{align}
Using these we write the following coupled integral equations for the
expectation values of $\hat{n}_{L}$, $\hat{n}_{R}$, $\tilde{s}$, $\tilde
{s}^{\dag}$,
\begin{align}
\left\langle \hat{n}_{L}\right\rangle _{t}-\left\langle \hat{n}_{L}%
\right\rangle _{0} &  =-\frac{i}{\hbar}\int_{0}^{t}dt^{\prime}\left[
T_{c}\left(  \left\langle \tilde{s}\left(  t\right)  \right\rangle
-\left\langle \tilde{s}^{\dag}\left(  t\right)  \right\rangle \right)
\right.  \nonumber\\
&  \left.  +\Gamma_{L}\left\langle n_{L}\right\rangle -\Gamma_{L}%
f_{S}\ \right]
\label{nL}%
\end{align}%
\begin{align}
\left\langle \hat{n}_{R}\right\rangle _{t}-\left\langle \hat{n}_{R}%
\right\rangle _{0} &  =\frac{i}{\hbar}\int_{0}^{t}dt^{\prime}\left[
T_{c}\left(  \left\langle \tilde{s}\left(  t\right)  \right\rangle
-\left\langle \tilde{s}^{\dag}\left(  t\right)  \right\rangle \right)
\right.  \nonumber\\
&  \left.  +\Gamma_{R}\left\langle n_{R}\right\rangle -\Gamma_{R}%
f_{D} \right]
\label{nR}%
\end{align}%
\begin{align}
\left\langle \tilde{s}\right\rangle _{t}-\left\langle \tilde{s}\right\rangle
_{0} &  =-\frac{i}{\hbar}\int_{0}^{t}dt^{\prime}e^{i\epsilon\left(
t-t^{\prime}\right)  }T_{c}\nonumber\\
&  \times\left\{  \left\langle n_{L}X_{t}X_{t^{\prime}}^{\dag}\right\rangle
_{t^{\prime}}-\left\langle n_{R}X_{t^{\prime}}^{\dag}X_{t}\right\rangle
_{t^{\prime}}\right\}  \nonumber\\
&  +\frac{\left(  \Gamma_{L}+\Gamma_{R}\right)}{2}  \int_{0}^{t}dt^{\prime}%
e^{i\epsilon\left(  t-t^{\prime}\right)  }\left\langle \tilde{s}\left(
t^{\prime}\right)  X_{t^{\prime}}^{\dag}X_{t}\right\rangle
\label{s}
\end{align}%
\begin{align}
\left\langle \tilde{s}^{\dag}\right\rangle _{t}-\left\langle \tilde{s}^{\dag
}\right\rangle _{0} &  =-\frac{i}{\hbar}\int_{0}^{t}dt^{\prime}e^{-i\epsilon
\left(  t-t^{\prime}\right)  }T_{c}\nonumber\\
&  \times\left\{  \left\langle n_{R}X_{t}^{\dag}X_{t^{\prime}}\right\rangle
-\left\langle n_{L}X_{t^{\prime}}X_{t}^{\dag}\right\rangle \right\}
\nonumber\\
&  +\frac{\left(  \Gamma_{L}+\Gamma_{R}\right)}{2}  \int_{0}^{t}dt^{\prime}%
e^{-i\epsilon\left(  t-t^{\prime}\right)  }\left\langle X_{t^{\prime}}%
X_{t}^{\dag}\tilde{s}^{\dag}\left(  t^{\prime}\right)  \right\rangle.
\label{s_dagger}%
\end{align}
This is still exact, but intractable. To make progress one assumes that the molecule operators and the
oscillator ones are separable, so $\left\langle n_{L}X_{t}X_{t^{\prime}}%
^{\dag}\right\rangle _{t^{\prime}}$ can be written as
\begin{align}
\left\langle n_{L}X_{t}X_{t^{\prime}}^{\dag}\right\rangle _{t^{\prime}} &
=\left\langle n_{L}\right\rangle \left(  1-e^{-\beta\omega_{B}}\right)
\left\langle e^{-n\beta\omega_{B}}X_{t}X_{t^{\prime}}^{\dag}\right\rangle
,\nonumber\\
&  \equiv\left\langle n_{L}\right\rangle \mathcal{F},
\end{align}
with $\beta=k_{B}T$. We have also assumed that the vibrational mode is in
equilibrium due to contact with a thermal bath at temperature $T$. The other
properties of the mode are ensconced in the two-time boson correlation
functions $\mathcal{F}$. In the appendix we give an analytical form for
$\mathcal{F}$ in the limit of an undamped single mode. Then, one can in
principle solve the above equations of motion for arbitrary molecule-vibron
coupling strength (which includes non-Markovian properties of this interaction).

The polaron transformation method is powerful in that it allows us to gain
simple analytical forms in some cases, and in general \textquotedblleft
plug-in" arbitrary bath correlation functions for $\mathcal{F}$ although we
need to be mindful of the regime where the method is valid. Some works specify
that it gives qualitatively correct results for small $T_{c}$, while, a comparison to the NIBA \cite{LeggettRMP} suggests it is very accurate for $\epsilon=0$ or $\epsilon\gg T_c, g$. For the case of a single-mode, which we consider here, understanding when the results are
accurate can be challenging, as we will discuss in the next section.  Generally, we do find them to be qualitatively correct, though additional back-action effects arise in the exact numerical treatments.

Finally, before showing explicit results, we make an additional assumption.  The above equations allow for double electron occupation
of the dimer molecule.  Allowing such occupation is interesting, but prevents us from obtaining analytical results.  Thus, as mentioned earlier, we impose another condition equivalent to the notion of Coulomb blockade in quantum dots.  This is valid if the molecule is small and the bias not too large (i.e., this Coulomb repulsion is the largest energy scale).  In this case only one electron can inhabit the dimer at a time and sequential tunneling occurs.

Rather than explicitly include
this effect with a non-linear repulsion against double-site occupation in the Hamiltonian, we can simply replace the fermionic operators with projectors onto the single electron states \beq n_L&=&\ket{L}\bra{L},\quad\quad n_R=\ket{R}\bra{R},\eeq with \beq n_L+n_R+\ket{0}\bra{0}=1,\eeq where $\ket{0}\bra{0}$ is the no-electron state.  Similarly, $s=\ket{L}\bra{R}$, $s^{\dagger}=\ket{R}\bra{L}$.  The combinations of operators in the Lindblad terms describing the coupling to the leads and  Equations [\ref{nL}-\ref{s_dagger}] are subtly altered by this change, to preserve the reduced Hilbert space.   We do not give explicit details here as the derivation is equivalent to the previously studied model of a double quantum dot coupled to a phonon bath\cite{STB04,Brandes05}, except where the single-mode-correlation function is replaced by the spectral function for a particular bath model.  For completeness we give the altered equations of motion in the appendix.  It is these equations of motion, under this additional assumption, which give us the results in the next section.

\section{Results}

\subsection{Current}

The expectation value of the current from the source to the left molecule is
$I_{L}\left(  t\right)  =-e\Gamma_{L}\left(  1-f_{S}\right)  \left\langle
n_{L}\right\rangle $, from the right molecule to the drain is $I_{R}\left(
t\right)  =-e\Gamma_{R}\left(  1-f_{D}\right)  \left\langle n_{R}%
\right\rangle $, and from the left molecule to the right molecule is
$I_{L}\left(  t\right)  =-eT_{c}\left(  \left\langle \tilde{s}\left(
t\right)  \right\rangle -\left\langle \tilde{s}^{\dag}\left(  t\right)
\right\rangle \right)  $. In the steady-state these three currents are the
same, and can easily be evaluated from the expectation values of the molecule
operators descried above.

The most straightforward method to solve such equations is to explicitly Laplace
transform the equations of motion first and then find the coefficient of the
Laplace parameter $1/z$ as $z\rightarrow0$ in the Laurent expansion of the
expectation values\cite{STB04}. We can also use the full $z$ dependent forms
to find the current-noise spectrum, as outlined in the next subsection. As an
example, we explicitly solve here for the current using the $\mathcal{F}$
function described in the appendix. As discussed earlier, this allows us to investigate the unusual
limit of strong coupling of the transport process to an undamped single mode.
Note, in all of the following results, we set $\hbar=1$, $e=1$ and typically $\omega
_{m}=1$, apart from in the noise power results.

\begin{figure}[pth]
\begin{center}
\includegraphics[width=\columnwidth] {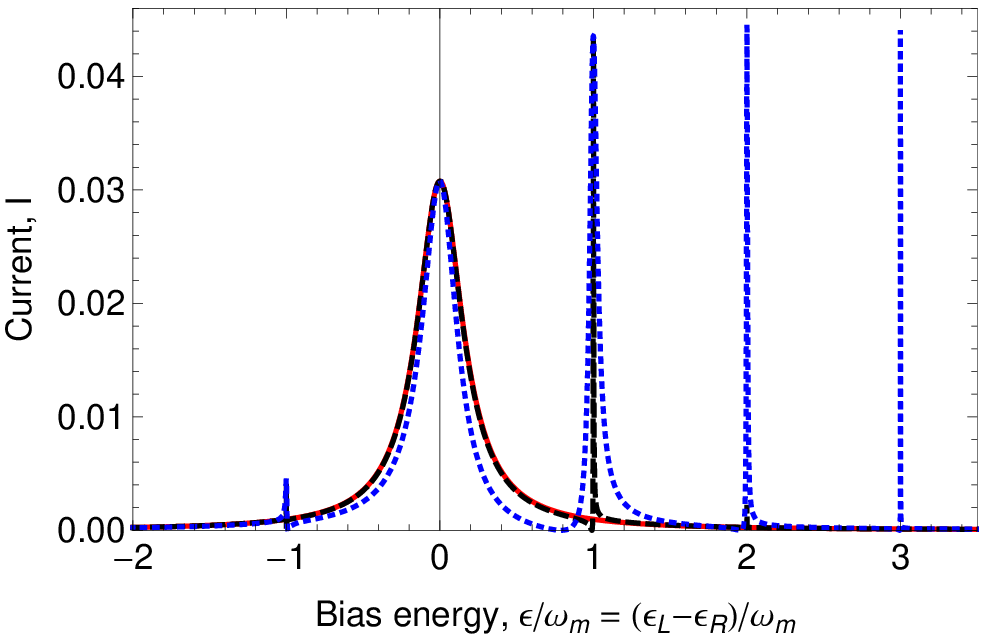}
\includegraphics[width=\columnwidth] {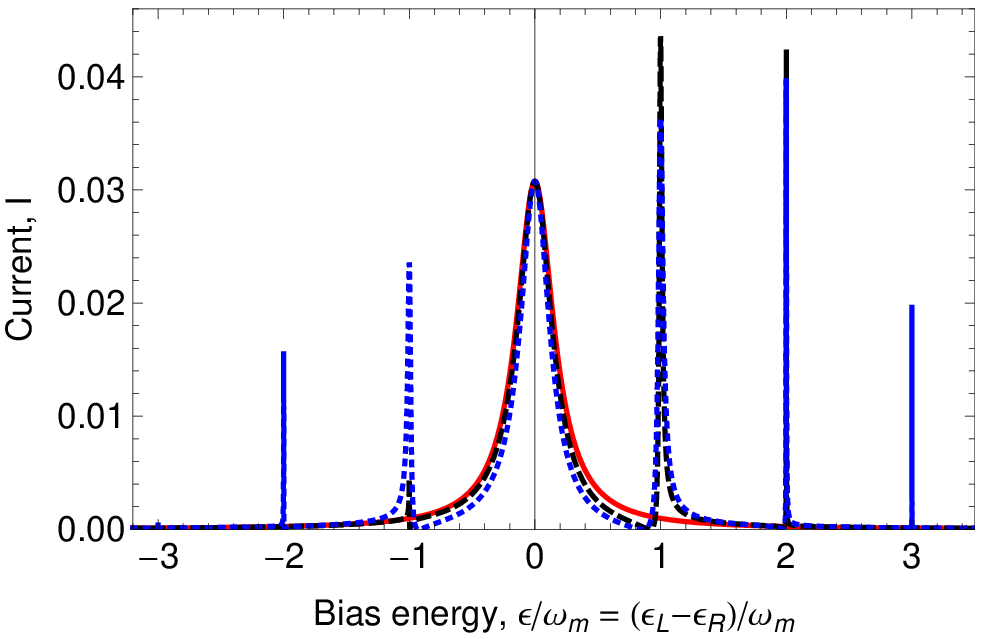}
\end{center}
\caption{(Color online) (Top) The current $I_{R}/e$, as given by the polaron
transformed results, through the molecule as a function of molecular energy
bias $\epsilon/\omega_m = (\epsilon_L-\epsilon_R)/\omega_m$. The solid line is the uncoupled case, while the dashed lines
are for different magnitudes of the coupling to the vibrational mode:
$z=g/\omega_{m}$, ($z=0.1$ black- and $z=0.5$ blue-dashed lines, with a thermal occupation of the
vibrational mode of $N=0.05$). Under this polaron transformation assumption, and for a
completely undamped vibrational mode, we see extremely strong vibrational
emission resonances in the current. (Bottom) The current $I_{R}/e$ for no
coupling (solid line) and for coupling $z=g/\omega_{m}=0.3$, and for several
different choices of the thermal occupation of the vibrational mode $N=0.01$ black- and $N=1$ blue-dashed lines. As we raise the thermal temperature of the mode, absorbtion resonances
occur, and the linewidth of the emission resonances becomes broadened.  In both figures we set $\Gamma_{L}=\Gamma_{R}=0.1$, $T_{c}=0.1$, and $\omega_m=1$.
 }%
\label{IvsEpsilon1}%
\end{figure}

The top figure in Fig.\ \ref{IvsEpsilon1} shows the current as a function of
the energy difference between the left and right molecules (scaled with the
resonant frequency of the vibrational mode) for various vibrational mode
coupling strengths ($z=0.1$ black- and $z=0.5$ blue-dashed lines) and without
the mode coupling (solid red line) with a thermal occupation of the
vibrational mode of $N=0.05$. The bottom figure in Fig.\ \ref{IvsEpsilon1}
shows the current for different temperature ($N=0.01$ black- and $N=1$ blue-
dashed lines) while keeping the coupling strength constant at $z=0.3$.

\begin{figure}[pth]
\begin{center}
\includegraphics[width=0.8\columnwidth] {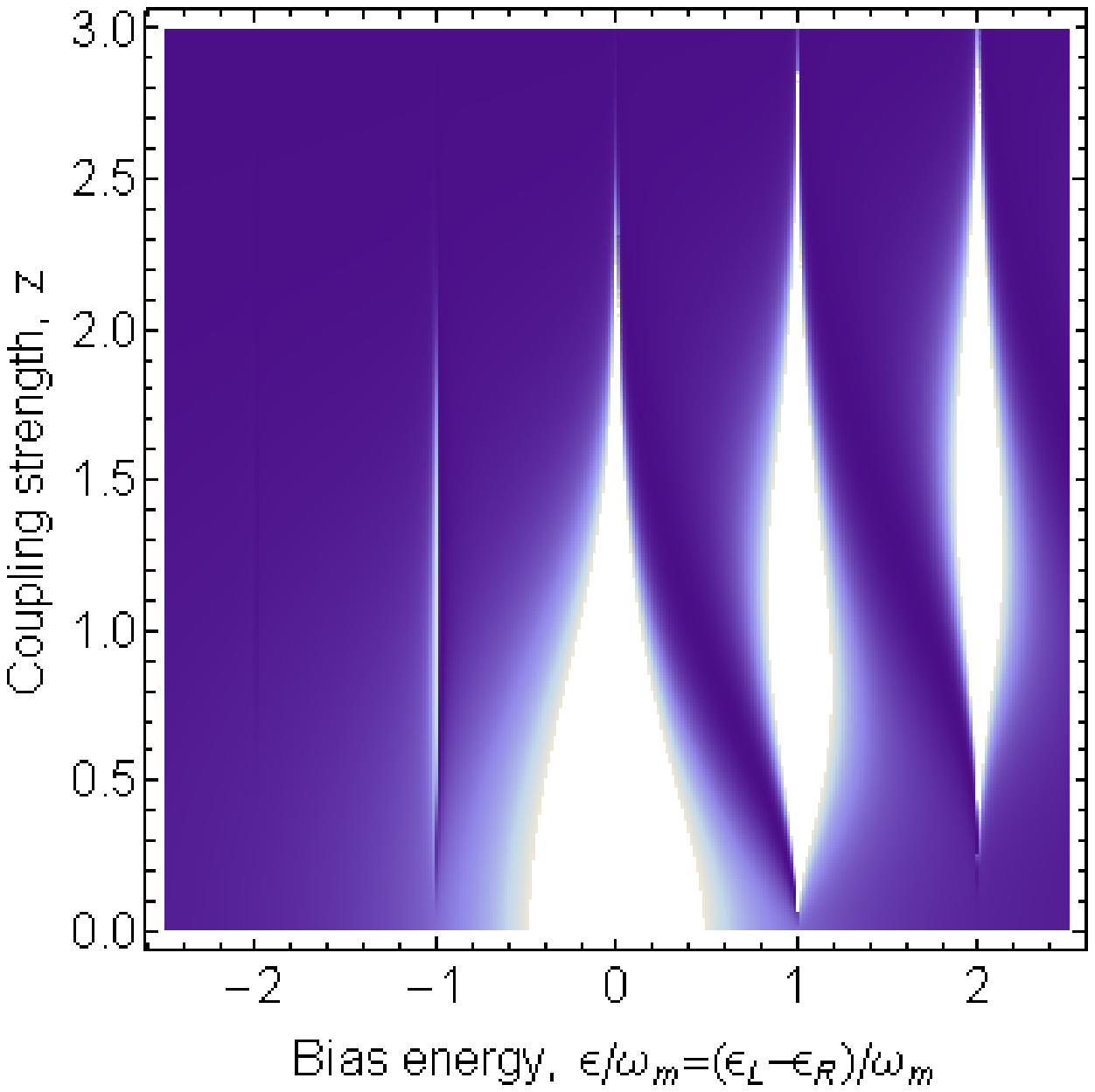}
\includegraphics[width=0.8\columnwidth] {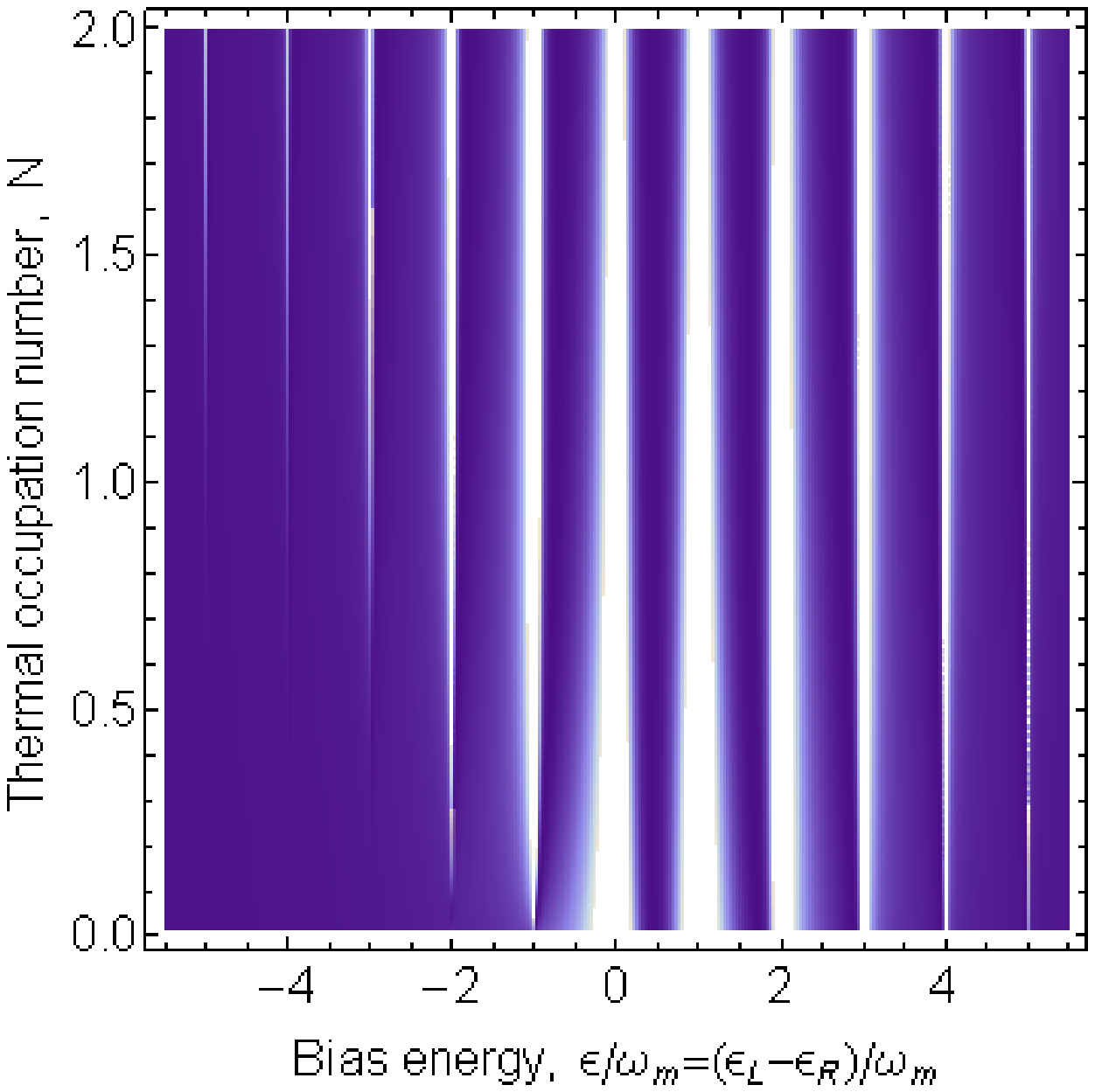}
\end{center}
\caption{(color online) (Top) The current $I_{R}/e$ through the molecule as a
function of the molecular energy bias $\epsilon/\omega_m$ and vibronic coupling strength
$z=g/\omega_{m}$, with thermal occupation $N=0.05$. (Bottom) The current $I_{R}/e$ for coupling $z=g/\omega_{m}
= 1$ as a function of $\epsilon$ and thermal occupation $N$. In both figures
white is a large current and purple is zero current, and these results are
from the polaron treatment, and again for $\Gamma_{L}=\Gamma_{R}=0.1$, $T_{c}=0.1$, $\omega_m=1$.}%
\label{figd}%
\end{figure}

When $\epsilon$ is much larger than the tunnel coupling, the electrons cannot
tunnel between left and right molecules and the current approaches zero.
With the vibrational mode coupling \textquotedblleft on\textquotedblright, the
current peaks when the energy bias $\epsilon$ equals the vibrational mode
resonant frequency and its multiples (as also observed in the case of quantum
dots coupled to a single phonon mode\cite{lambert1}). Positive
$\epsilon$ spikes correspond to phonon-emission-assisted transport and
negative $\epsilon$ spikes correspond to phonon-absorption. In figure
\ref{figd} we see that increasing the coupling strength narrows the lower
emission peaks and broadens the higher peaks. The bottom figure in
Fig.\ \ref{figd} shows how raising the temperature broadens and raises the
absorption peaks. In addition, between peaks there is a strong suppression of
the current, akin to the Franck-Condon blockade. If we directly observe, around the first resonance, the
steady state occupations of the various states we find that this
blockade effect strongly localizes the electrons in the left molecule.

The amplitudes of the absorption peaks strongly depend on temperature.
Expanding the expression for the current around the first absorption resonance
we find
\begin{equation}
I_{max}^{\mathrm{abs}}=\left(  \Gamma_{L}^{-1}+\frac{\Gamma_{R}}{4T_{c}^{2}%
}+\frac{1+2N}{\Gamma_{R}(1+N)}\right)  ^{-1}.%
\end{equation}
In contrast, the height of the emission peaks is only weakly dependent on the
thermal occupation, and, exactly at the resonant point, the maximum is not dependant on the coupling strength or the order, $\epsilon=n\omega$, of the resonance. For low temperature the
height of resonance is proportional to the zero-bias current
\begin{align}
I_{max}^{\mathrm{emi}}  & =I_{\epsilon=0}(1+I_{\epsilon=0}),\\
I_{\epsilon=0}  & =\frac{4T_{c}^{2}}{\Gamma_{R}+4T_{c}^{2}(2\Gamma_{R}%
^{-1}+\Gamma_{L}^{-1})}.
\end{align}
However the width around the resonance is strongly dependent on both the coupling strength and the
order $n\omega$ of the resonance peak. In addition, the properties of each
resonance peak are almost entirely defined by the appropriate term retained in
the sum in Eq. (\ref{sumeq}). Thus, overall, we can say that in-between
resonances there is a strong suppression of the current due to the strong
coupling to the vibrational mode, and exactly on resonance (for emission) the
transport channel is completely transparent. This represents a unique kind of
vibrationally-mediated transistor, where the transport can be exquisitely
sensitive to small changes in the energies of the electronic states in the dimer.

The fact that the height of the peaks is independent of the coupling strength
is surprising. Under exactly what regime are these results valid? Recall that
we made two approximations in the derivation; we assumed that the electronic
and bosonic operators were separable, and that the bosonic correlation
functions were entirely described by their thermal equilibrium state, so that
\begin{equation}
\left\langle n_{L}X_{t}X_{t^{\prime}}^{\dag}\right\rangle _{t^{\prime}%
}=\left\langle n_{L}\right\rangle \left(  1-e^{-\beta\omega_{B}}\right)
\left\langle e^{-n\beta\omega_{B}}X_{t}X_{t^{\prime}}^{\dag}\right\rangle .
\end{equation}
This implies a neglect of the back-action onto the single-mode dynamics.
In addition, we assumed that the dynamics of the bosonic system were not damped
(though this can be included via numerical evaluations of the correlation
function). These two approximations can be interpreted in the following way;
that we have coupled the electron transport to a semi-\emph{classical}
oscillator that is capable of emitting and absorbing energy at very specific
frequencies. 
The fact that we neglect back-action does indicate there should be a
quantitative difference to arise from the exact solution. In the final section we will compare these results to two exact numerical models.
In addition, as mentioned earlier, the set of
approximations made here is, at least in the many-mode case, formally equivalent to the non-interacting blip
approximation \cite{LeggettRMP,Dekker86}. 

\subsection{Noise spectrum}

\begin{figure}[pth]
\begin{center}\includegraphics[width=\columnwidth] {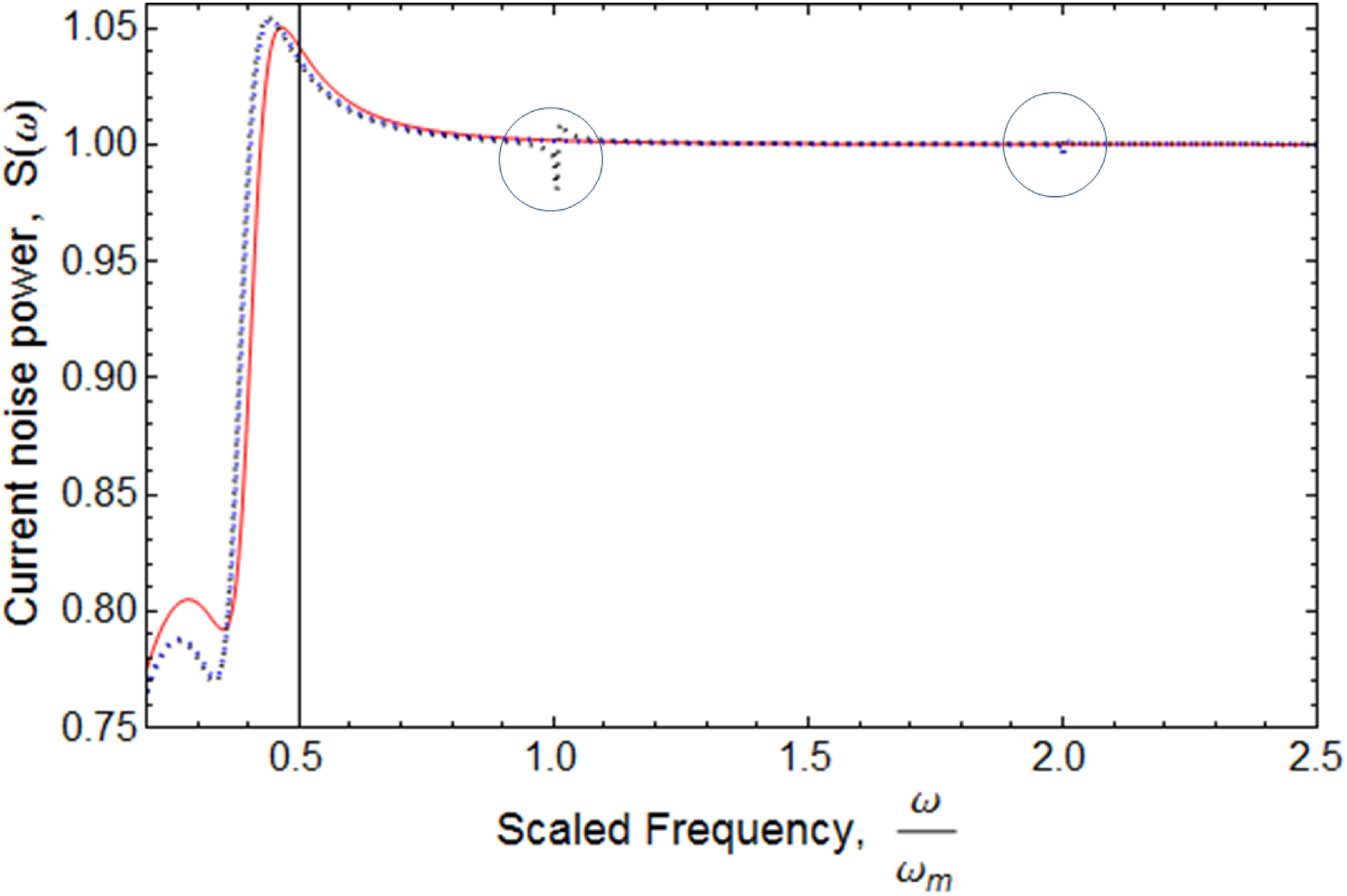}
\end{center}
\caption{(Color online) The current noise power Fano factor $F_R\left(  \omega\right) = S_{R}\left(  \omega\right)/2eI_R  $, for
$\Gamma_{L}=\Gamma_{R}=0.1$, $T_{c}=0.1$, $z=0,0.3$, and $\omega_{m}=1,2$, and
$N=0.05$. There are two resonances $\omega=2T_{c}$ and $\omega
=\omega_{m}$, as well a an additional higher resonance at $2\omega_{m}$. The latter two resonances are indicated by the two blue circles.}%
\label{fignoise}%
\end{figure}

The current-noise power spectral density is a useful and easy way, with this formalism,
to gain some insight on the transient dynamics.  The current noise power has three  components: particle currents
through left molecule $S_{L}\left(  \omega\right)  $, the right molecule $S_{R}\left(  \omega\right)  $ and the charge noise spectrum
$S_{CN}\left(  \omega\right)  $ \cite{AB04}:
\begin{equation}
S\left(  \omega\right)  =\alpha^{2}S_{L}\left(  \omega\right)  +\beta^{2}%
S_{R}\left(  \omega\right)  -\alpha\beta\omega^{2}S_{CN}\left(  \omega\right)
.
\end{equation}
The coefficients $\alpha$ and $\beta$, with $\alpha+\beta=1$, depend on the
capacitances between junctions and molecule.
Here we focus on the contribution from the right particle current, which can
dominate if $\beta\gg\alpha$. This contribution can be obtained from the
MacDonald formula \cite{AB04,Mac48} and%
\begin{align}
S_{R}\left(  \omega\right)   &  =2\omega e^{2}\int_{0}^{\infty}dt\text{ }%
\sin\left(  \omega t\right)  \frac{d}{dt}\left[  \left\langle n^{2}\left(
t\right)  \right\rangle -\left(  t\left\langle I\right\rangle ^{2}\right)
\right]  ,\nonumber\\
&  =2eI\left\{  \mathcal{L}\left\langle \hat{n}_{i}\left(  t\right)  ,\hat
{n}_{j}\left(  t+\tau\right)  _{i\omega}\right\rangle \right.  \nonumber\\
&  \left.  +\mathcal{L}\left\langle \hat{n}_{i}\left(  t\right)  ,\hat{n}%
_{j}\left(  t+\tau\right)  \right\rangle _{-i\omega}\right\}  .
\end{align}
The steady state of $S_{R}\left(  \omega\right)  $ can be calculated as%
\begin{equation}
S_{R}\left(  \omega\right)  =2eI\left\{  1+\Gamma_{R}\left[  \hat{n}%
_{R}\left(  -i\omega\right)  +\hat{n}_{R}\left(  i\omega\right)  \right]
\right\}
\end{equation}
The frequency dependent $\hat{n}_{R}\left(  i\omega\right)  $ is again easily
extracted from the Laplace transform of Eqs.~[\ref{nL}] and [\ref{nR}].
However, as noted in \cite{Brandes05} the validity of the noise spectrum
formulation with the polaron transformation treatment is ambiguous can easily produce
non-physical results. Figure \ref{fignoise} show
the spectral density with both no coupling (red line) and weak coupling for
two different vibrational frequencies $\omega_{m}$ ($\omega_m=1$, black- and
$\omega_m=2$, blue- dashed lines) for $T_{c}=0.2$, $\Gamma_{L}=\Gamma_{R}=0.1$,
and $N=0.05$. There is a clear large resonance from the bare Rabi oscillations of
the electron tunneling between molecules at $\omega=2T_{c}=0.4$, as well
additional resonances at $\omega=\omega_{m}$.

\section{Numerical comparison: exact solution and the Hierarchy equation}

For small thermal occupation $N$, reasonable coupling strength $z$ and by
including damping of the vibrational mode (or an effective multi-mode
environment) we can solve the molecule-vibration coupling exactly with a
variety of different numerical approaches. Here we use the Hierarchy equation
of motion for a Lorentz bath spectral density recently introduced by Ma
\textit{et al} \cite{ma}. Essentially we take our original equation of motion,
but write the interaction with the vibrational mode as a bath of oscillators.
We then assume that this bath has a Lorentz spectrum,%
\begin{equation}
J(\omega)=\frac{1}{\pi}\frac{\lambda\gamma}{(\omega-\omega_{m})^{2}+\gamma
^{2}},
\end{equation}
where $\lambda$ is equal to the square of the molecule-vibron coupling
strength $\lambda=g^{2}$, $\omega_{m}$ is the fundamental frequency of
the vibronic excitation, and $\gamma$ is the broadening or damping of the
vibronic mode. We then follow the steps in Ref.\ [\onlinecite{ma}] to reach the
following hierarchical equation of motion (again setting $\hbar=1$),%
\begin{align}
\frac{\partial}{\partial t}\rho_{\vec{n}}(t) &  =-(\mathcal{L}+n\cdot\mu
)\rho_{\vec{n}}(t)\nonumber\\
&  -i\sum_{k=1}^{2}V^{\times}\rho_{\vec{n}+\vec{e}_{k}}(t)\nonumber\\
&  -i\frac{\lambda}{2}\sum_{k=1}^{2}n_{k}\left[  V^{\times}+(-1)^{k}%
V^{0}\right]  \rho_{\vec{n}-\vec{e}_{k}}(t).
\end{align}
 The superoperator notation introduced in this equation is $V^{\times}\centerdot=[V,\centerdot]$ and $V^{0}\centerdot=\{V,\centerdot\}$, where $V$ is the electronic operator which describes the coupling to the Lorentz bath, which in this case is $V=n_L - n_R$. Here, $\vec{n}$ is a two-dimensional index with positive integer
elements identifying both the true system density matrix $\rho_{(0,0)}$ and
the auxiliary operators which encode the effect of the bath $\rho_{(>0,>0)}$.
The other vectors are defined as $\vec{e}_{1}=(1,0)$, $\vec{e}_{2}=(0,1)$ and
$\vec{\mu}=(\gamma-i\omega_{0},\gamma+i\omega_{0})$.
 $\mathcal{L}$ is the super-operator describing the coherent and
incoherent dynamics of the molecule transport.  Note that again here we assume a large double-occupation repulsion (Coulomb blockade), so that the electronic and transport properties are simply described by the single electron occupation projectors, $n_L=\ket{L}\bra{L}$,  $n_R=\ket{R}\bra{R}$, $s=\ket{L}\bra{R}$ and,
\beq
\mathcal{L}[\centerdot]&=& -i \left[\frac{\epsilon}{2}(n_L - n_R) + T_c(s+s^{\dagger}), \centerdot\right] \nonumber\\
&-&\frac{\Gamma_L}{2} D\left[S_{L}\right] - \frac{\Gamma_R}{2} D\left[S_R^{\dagger}\right].
\eeq
where $S_L=\ket{0}\bra{L}$, $S_R=\ket{0}\bra{R}$, and $\mathcal{D}$ is as defined in Eqs. (\ref{D}-\ref{D2}).

The tier of the hierarchy
in this method must be truncated at some reasonable value, $\vec{n}%
=(N_{c},N_{c})$. Typically this is done by increasing $N_{c}$ until
convergence is found. If $\lambda$ is increased or $\gamma$ is decreased the
$N_{c}$ necessary for convergence rises (representing the increased occupation
of the bath, and the increase in associated degrees of freedom). The
hierarchical equations of motion are solved by setting $\rho_{(0,0)}$ to the
desired molecule initial state, setting all auxiliary density matrices to zero $\rho_{(>0,>0})=0$, and employing direct
numerical integration. Here we are primarily interested in steady-state
results so we integrate until the system probabilities become stationary. The
steady state can also be found by finding the null vector of the matrix
defining these equations of motion.

\begin{figure}[pth]
\begin{center}
\includegraphics[width=\columnwidth] {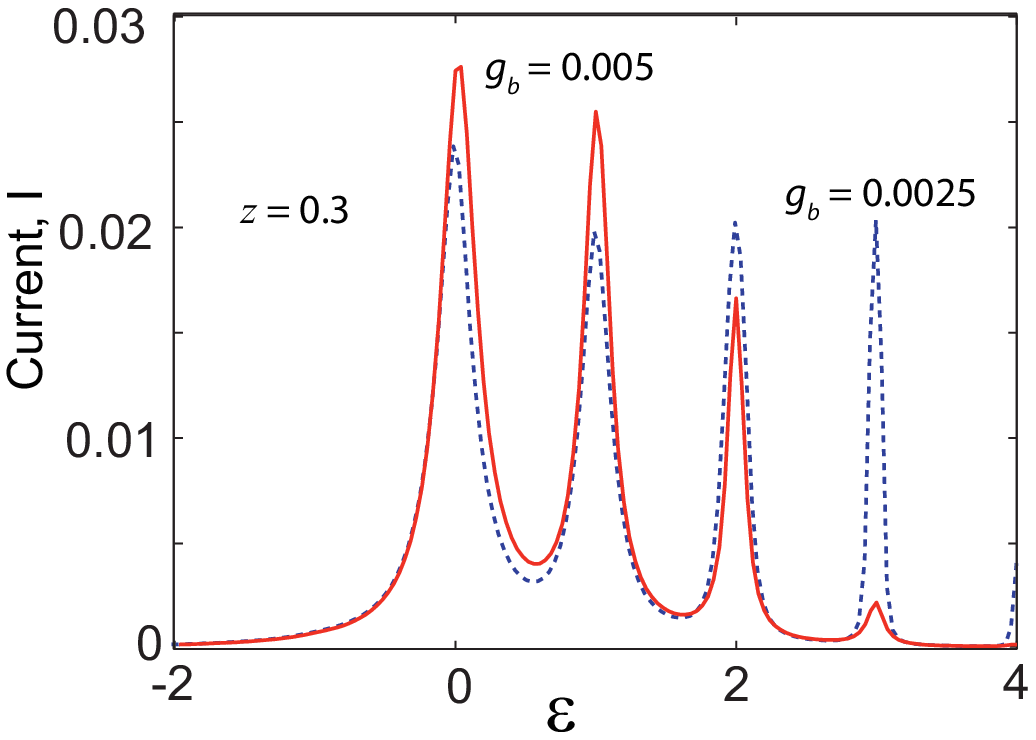}
\includegraphics[width=\columnwidth] {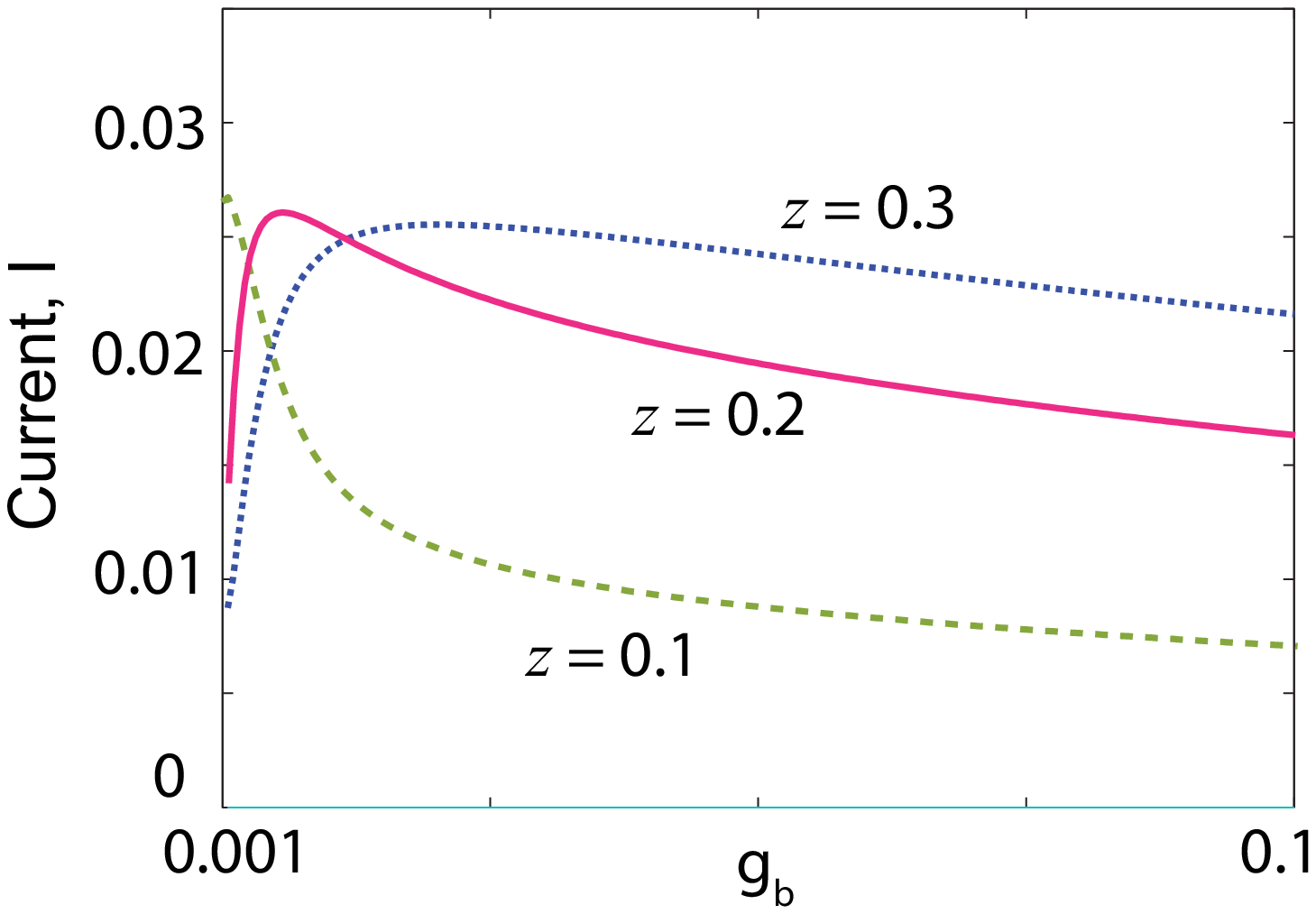}
\end{center}
\caption{(Top) The current  $I_{R}/e$  as given by the exact numerical treatment for $z=0.3$.
The red solid curve is for $\gamma_{b}=0.005$ and the dashed blue curve is for
$\gamma_{b}=0.0025$. We find that as we reduce the vibrational mode damping
current qualitatively behaves like the polaron treatment results. However this
is an additional overall reduction in magnitude not predicted by the polaron
results. (Bottom) The current $I_R/e$ as a function of damping for $\epsilon=\omega_m$, for three different values of the
coupling strength. We see that the three curves have a similar maxima, but are suppressed when $\gamma_b$ is reduced (the $z=0.1$ curve is also suppressed for $\gamma_b < 0.001$).}%
\label{Ivsg}%
\end{figure}

To check the validity of this method we also directly diagonalize the equation
of motion (without the polaron transform, and with an additional Lindblad
damping of the vibrational mode with rate $\gamma_{b}=\gamma/2$) in the Fock
basis, and integrate the resulting coupled equations of motion. We find that,
given convergence, the results for the two numerical methods
in the steady-state match exactly. Comparing to the polaron results we see
some similarities and some differences. In the top figure of
\ref{Ivsg} we see that the numerics, like the polaron treatment,
predict equal height emission peaks for the different resonances as
$\gamma_{b}$ becomes small. However as $\gamma_{b}$ is decreased further there
is also a reduction in the overall current magnitude. This is an additional
localization of the electron due to the strong coupling to the vibration not
captured by the polaron treatment.  This phenomena is more clearly shown in the bottom figure of \ref{Ivsg} where
we show the current versus the vibrational mode damping at the first resonance peak, $\epsilon=\omega_m$.  We see that for all
coupling strengths there is a common maxima following by a suppression of the current as $\gamma_b$ is reduced.  This indicates that, for the purpose of current enhancement, there is an optimal finite vibrational mode damping.

\section{Conclusion}

In conclusion, we have analyzed the properties of coherent electron transport through a
dimer that is strongly coupled to a vibrational mode. When the energy
difference between the electronic states in the two sites of the dimer is larger than the tunneling coupling, the current
is suppressed. On the other hand, a semi-analytical polaron treatment suggests, when this energy difference is near the vibrational resonant frequency, an
enhancement of the electron transport appears as extremely sharp and
equal-height current resonances.
This strong coupling also induces Frank-Condon blockade, suppressing the current
between the resonances. The vibrational mode temperature determines the nature of the phonon-assisted
transport. At low temperature, phonon-emission-assisted electron
transport is the dominant feature. As the temperature is increased, the phonon-absorption-assisted
electron transport also starts to play a role.

We also examined two exact numerical treatments of the same system. These two treatments were found to be identical in the steady-state,
and qualitatively agreed with the polaron treatment. In addition, these results suggest that an additional localization or current suppression mechanism appears for very low vibrational mode damping, and thus there exists an optimal finite damping.

Our investigation is primarily motivated by studies of electron transport
through single molecules. However, our analysis is also applicable to
artificial systems, in particular quantum dots in carbon nanotubes, where
strong coupling can occur. The connection to large systems like doubly-clamped
semiconductor beams and cantilevers is more tenuous, as in those cases we
expect the coupling to be weak and thermal phonon effects to dominate the
transport properties.


\section{Acknowledgements}

DHS thanks RIKEN for hospitality and a stimulating environment while this work was done.
FN acknowledges partial support from the ARO, JSPS-RFBR contract No.
09-02-92114, MEXT Kakenhi on Quantum Cybernetics, and the JSPS-FIRST Program.

\section{Appendices}

\subsection{Boson correlation functions}

Here we show how to derive the analytical form for the correlation function of
a single undamped oscillator mode. This is equivalent to a recent derivation
given in [\onlinecite{esposito}]. We start with a similarity
transformation%
\begin{align}
e^{-M}f\left(  \hat{A}\right)  e^{M} &  =e^{-M}\left(  \sum_{n=0}a_{n}\hat
{A}\right)  e^{M}\\
&  =\sum_{n=0}a_{n}e^{-M}\hat{A}e^{M}=\sum_{n=0}a_{n}\tilde{A}=f\left(
\tilde{A}\right)  \nonumber
\end{align}
which suggests that one needs to consider only the transformation on each
operator separately. Utilizing the interaction picture of the boson operators
and their commutation%
\begin{equation}
\left[  -z\hat{b}^{\dag}e^{i\omega t},z\hat{b}e^{-i\omega t}\right]  =z^{2},
\end{equation}
$X\left(  t\right)  $ and $X^{\dag}\left(  t\right)  $ operators become
\begin{align}
X\left(  t\right)   &  =\exp\left[-z\left(  \hat{b}^{\dag}e^{i\omega t}-\hat
{b}e^{-i\omega t}\right)  \right]\nonumber\\
&  =\exp\left(-z\hat{b}^{\dag}e^{i\omega t}\right)\exp\left(z\hat{b}e^{-i\omega t}\right)\nonumber\\
& \times \exp\left(-\frac{1}{2}\left[  z\hat{b}^{\dag}e^{-i\omega t},z\hat{b}e^{-i\omega t}\right]\right)\nonumber\\
&  =\exp{-z\hat{b}^{\dag}e^{i\omega t}}\exp{z\hat{b}e^{-i\omega t}}e^{-\frac{1}%
{2}z^{2}}%
\end{align}%
\begin{equation}
X^{\dag}\left(  t\right)  =e^{-\frac{1}{2}z^{2}}\exp\left(z\hat{b}^{\dag}e^{i\omega
t}\right)\exp\left(-z\hat{b}e^{-i\omega t}\right)%
\end{equation}
then, $X_{t}X_{t^{\prime}}^{\dag}$ becomes
\begin{equation}
X_{t}X_{t^{\prime}}^{\dag}=e^{-z^{2}}e^{-z\hat{b}^{\dag}e^{i\omega t}}\left(
e^{z\hat{b}e^{-i\omega t}}e^{z\hat{b}^{\dag}e^{i\omega t^{\prime}}}\right)
e^{-z\hat{b}e^{-i\omega t^{\prime}}}%
\end{equation}
Putting the operators in normal order by using
\begin{align}
e^{iH_{T}t/\hbar}\mathcal{O}e^{-iH_{T}t/\hbar} &  =\mathcal{O}+\left[
\frac{i}{\hbar}tH_{T},\mathcal{O}\right]  \\
&  +\frac{1}{2!}\left[  \frac{i}{\hbar}tH_{T},\left[  \frac{i}{\hbar}%
tH_{T},\mathcal{O}\right]  \right]  +\ldots.\nonumber
\end{align}
and after noticing $e^{-M}f\left(  \hat{A}\right)  e^{M}=e^{-z\hat{b}_{t^{\prime}%
}^{\dag}}e^{z\hat{b}_{t}}e^{z\hat{b}_{t^{\prime}}^{\dag}}$, with $f\left(
\hat{A}\right)  =e^{z\hat{b}_{t}}$, $\hat{A}=ze^{-i\omega t}\hat{b}$,
\begin{equation}
e^{-z\hat{b}_{t^{\prime}}^{\dag}}\left(  \hat{b}e^{-i\omega t}\right)
e^{z\hat{b}_{t^{\prime}}^{\dag}}=e^{-i\omega t}\left(  b+ze^{i\omega t\prime
}\right)  \nonumber
\end{equation}%
\begin{equation}
\Rightarrow e^{-z\hat{b}_{t^{\prime}}^{\dag}}e^{z\hat{b}_{t}}e^{z\hat
{b}_{t^{\prime}}^{\dag}}=e^{ze^{-i\omega t}b}\exp\left[  z^{2}e^{-i\omega
\left(  t-t^{\prime}\right)  }\right]
\end{equation}
so that%
\begin{equation}
e^{z\hat{b}_{t}}e^{z\hat{b}_{t^{\prime}}^{\dag}}=\exp\left[  z^{2}%
e^{-i\omega\left(  t-t^{\prime}\right)  }\right]  e^{z\hat{b}^{\dag}e^{i\omega
t\prime}}e^{ze^{-i\omega t}b}.
\end{equation}
and we obtain%
\begin{align}
X_{t}X_{t^{\prime}}^{\dag} &  =\exp\left[  -z^{2}\left(  1-e^{-i\omega\left(
t-t^{\prime}\right)  }\right)  \right]  \exp\left[  z\hat{b}^{\dag}\left(
e^{i\omega t\prime}-e^{i\omega t}\right)  \right]  \nonumber\\
&  \times\exp\left[  -zb\left(  e^{-i\omega t^{\prime}}-e^{-i\omega t}\right)
\right]
\end{align}
The correlation $\mathcal{F}$ can be arranged in more a convenient form%
\begin{align}
\mathcal{F} &  =\left(  1-e^{-\beta\omega_{B}}\right)  \sum_{n=0}^{\infty
}\left\langle n\left\vert e^{-\beta n\omega_{B}}X_{t}X_{t^{\prime}}^{\dag
}\right\vert n\right\rangle \nonumber\\
&  =\left(  1-e^{-\beta\omega_{B}}\right)  e^{-z^{2}\left(  1-e^{-i\omega
_{B}\left(  t-t^{\prime}\right)  }\right)  }\nonumber\\
&  \times\sum_{n=0}^{\infty}\left\langle n\left\vert e^{-\beta n\omega_{B}%
}e^{z\hat{b}^{\dag}\left(  e^{i\omega t\prime}-e^{i\omega t}\right)
}e^{-zb\left(  e^{-i\omega t^{\prime}}-e^{-i\omega t}\right)  }\right\vert
n\right\rangle .\label{sumeq}%
\end{align}
Let $u=z\left(  e^{-i\omega t^{\prime}}-e^{-i\omega t}\right)  $, $u^{\ast
}=z\left(  e^{i\omega t^{\prime}}-e^{i\omega t}\right)  $, and expanding the
exponents in a power series and use the property of destruction operators will
condense the form to the Laguerre polynomial or order $n$
\begin{align}
\left\langle n\left\vert e^{u^{\ast}\hat{b}^{\dag}}e^{-u\hat{b}}\right\vert
n\right\rangle  &  =\sum_{l=0}^{n}\frac{\left(  -1\right)  ^{l}}{\left(
l!\right)  ^{2}}\frac{n!}{\left(  n-l\right)  !}\left(  \left\vert
u\right\vert ^{2}\right)  ^{l}\nonumber\\
&  =L_{n}\left(  \left\vert u\right\vert ^{2}\right)
\end{align}
Using one of its generating functions $\sum_{n=0}L_{n}\left(  \left\vert
u\right\vert ^{2}\right)  \xi^{n}=e^{-N\left\vert u\right\vert ^{2}}%
/(1-\xi)$ gives%
\begin{equation}
\mathcal{F}\left(  \tau\right)  =\exp\left\{  -z^{2}\left[  \left(  1+N\right)
\left(  1-e^{-i\omega\tau}\right)  +N\left(  1-e^{i\omega\tau}\right)
\right]  \right\}  \label{F_exp}%
\end{equation}
where $\tau=t-t^{\prime}$. This form is still not easy to evaluate in the
integral form of Eq.\ (\ref{s_dagger}). Therefore, we expand the exponential part
of Eq.\ (\ref{F_exp}) in terms of $\exp{\pm i\omega\tau}$%
\begin{equation}
\mathcal{F}\left(  \tau\right)  =e^{-z^{2}\left(  1+2N\right)  }\sum
_{s,p=0}^{\infty}\left(  z^{2}\right)  ^{s+p}\frac{N^{s}\left(  1+N\right)
^{p}}{s!p!}e^{i\left(  s-p\right)  \omega\tau}.
\end{equation}
Physically, the $s$\ terms account for absorption and $p$ for emission of
phonons\cite{esposito} with frequency $\omega$ . Writing as
\begin{align}
n &  =p-s=n_{\mathrm{emit}}-n_{\mathrm{ab}},\\
p &  =s+n,
\end{align}
we can rewrite \begin{widetext}
\begin{equation}
\mathcal{F}\left(  \tau\right)  =\sum_{n=-\infty}^{\infty}\exp{-in\omega\tau
}\exp{-z^{2}\left(  1+2N\right)  }\left(  \frac{1+N}{N}\right)  ^{\frac{n}{2}%
}I_{n}\left(  2z^{2}\sqrt{N\left(  1+N\right)  }\right)  ,
\end{equation}
\end{widetext}where $I_{n}$ is the modified Bessel function of the first kind.
This form can be numerically evaluated. Similarly, the other correlation
functions are%
\begin{align}
\left\langle X_{t^{\prime}}^{\dag}X_{t}\right\rangle  &  =\left\langle
X_{t^{\prime}}X_{t}^{\dag}\right\rangle =\mathcal{F}^{\ast}\left(
\tau\right)  ,\\
\left\langle X_{t}^{\dag}X_{t^{\prime}}\right\rangle  &  =\mathcal{F}\left(
\tau\right)  .
\end{align}

\subsection{Equations of motion for the Coulomb blockade limit}

To obtain the analytical and numerical results we show in the second half of this work we imposed an additional
condition on our model, that of strong Coulomb repulsion against double occupation of the dimer.  In this case we replace the operators with projectors onto a single electron basis, and obtain the following slightly altered equations of motion.
\begin{align}
\left\langle \hat{n}_{L}\right\rangle _{t}-\left\langle \hat{n}_{L}%
\right\rangle _{0} &  =-\frac{i}{\hbar}\int_{0}^{t}dt^{\prime}\left[
T_{c}\left(  \left\langle \tilde{s}\left(  t\right)  \right\rangle
-\left\langle \tilde{s}^{\dag}\left(  t\right)  \right\rangle \right)
\right.  \nonumber\\
&  \left.  +\Gamma_{L}\left(1-\left\langle n_{L}\right\rangle- \left\langle n_{R}\right\rangle\right) \right]
\label{nL2}%
\end{align}%
\begin{align}
\left\langle \hat{n}_{R}\right\rangle _{t}-\left\langle \hat{n}_{R}%
\right\rangle _{0} &  =\frac{i}{\hbar}\int_{0}^{t}dt^{\prime}\left[
T_{c}\left(  \left\langle \tilde{s}\left(  t\right)  \right\rangle
-\left\langle \tilde{s}^{\dag}\left(  t\right)  \right\rangle \right)
\right.  \nonumber\\
&  \left.  -\Gamma_{R}\left\langle n_{R}\right\rangle \right]
\label{nR2}%
\end{align}%
\begin{align}
\left\langle \tilde{s}\right\rangle _{t}-\left\langle \tilde{s}\right\rangle
_{0} &  =-\frac{i}{\hbar}\int_{0}^{t}dt^{\prime}e^{i\epsilon\left(
t-t^{\prime}\right)  }T_{c}\nonumber\\
&  \times\left\{  \left\langle n_{L}X_{t}X_{t^{\prime}}^{\dag}\right\rangle
_{t^{\prime}}-\left\langle n_{R}X_{t^{\prime}}^{\dag}X_{t}\right\rangle
_{t^{\prime}}\right\}  \nonumber\\
&  - \frac{\Gamma_{R}}{2}  \int_{0}^{t}dt^{\prime}%
e^{i\epsilon\left(  t-t^{\prime}\right)  }\left\langle \tilde{s}\left(
t^{\prime}\right)  X_{t^{\prime}}^{\dag}X_{t}\right\rangle
\label{s2}
\end{align}%
\begin{align}
\left\langle \tilde{s}^{\dag}\right\rangle _{t}-\left\langle \tilde{s}^{\dag
}\right\rangle _{0} &  =-\frac{i}{\hbar}\int_{0}^{t}dt^{\prime}e^{-i\epsilon
\left(  t-t^{\prime}\right)  }T_{c}\nonumber\\
&  \times\left\{  \left\langle n_{R}X_{t}^{\dag}X_{t^{\prime}}\right\rangle
-\left\langle n_{L}X_{t^{\prime}}X_{t}^{\dag}\right\rangle \right\}
\nonumber\\
&  -\frac{\Gamma_{R}}{2} \int_{0}^{t}dt^{\prime}%
e^{-i\epsilon\left(  t-t^{\prime}\right)  }\left\langle X_{t^{\prime}}%
X_{t}^{\dag}\tilde{s}^{\dag}\left(  t^{\prime}\right)  \right\rangle
\label{s_dagger2}%
\end{align}

\end{document}